\documentclass[conference]{IEEEtran}
\IEEEoverridecommandlockouts
\usepackage{cite}
\usepackage{amsmath,amssymb,amsfonts}
\usepackage{graphicx}
\usepackage{textcomp}
\usepackage{xcolor}
\def\BibTeX{{\rm B\kern-.05em{\sc i\kern-.025em b}\kern-.08em
    T\kern-.1667em\lower.7ex\hbox{E}\kern-.125emX}}

\usepackage[acronyms,nonumberlist,nopostdot,nomain,nogroupskip,acronymlists={hidden}]{glossaries}
\newacronym{3gpp}{3GPP}{3rd Generation Partnership Project}
\newacronym{4g}{4G}{4th generation}
\newacronym{5g}{5G}{5th generation}
\newacronym{6g}{6G}{6th generation}
\newacronym{5gc}{5GC}{5G Core}
\newacronym{adc}{ADC}{Analog to Digital Converter}
\newacronym{aerpaw}{AERPAW}{Aerial Experimentation and Research Platform for Advanced Wireless}
\newacronym{ai}{AI}{Artificial Intelligence}
\newacronym{aimd}{AIMD}{Additive Increase Multiplicative Decrease}
\newacronym{am}{AM}{Acknowledged Mode}
\newacronym{amc}{AMC}{Adaptive Modulation and Coding}
\newacronym{amf}{AMF}{Access and Mobility Management Function}
\newacronym{aops}{AOPS}{Adaptive Order Prediction Scheduling}
\newacronym{api}{API}{Application Programming Interface}
\newacronym{apn}{APN}{Access Point Name}
\newacronym{ap}{AP}{Application Protocol}
\newacronym{aqm}{AQM}{Active Queue Management}
\newacronym{ausf}{AUSF}{Authentication Server Function}
\newacronym{avc}{AVC}{Advanced Video Coding}
\newacronym{awgn}{AGWN}{Additive White Gaussian Noise}
\newacronym{balia}{BALIA}{Balanced Link Adaptation Algorithm}
\newacronym{bbu}{BBU}{Base Band Unit}
\newacronym{bdp}{BDP}{Bandwidth-Delay Product}
\newacronym{ber}{BER}{Bit Error Rate}
\newacronym{bf}{BF}{Beamforming}
\newacronym{bler}{BLER}{Block Error Rate}
\newacronym{brr}{BRR}{Bayesian Ridge Regressor}
\newacronym{bs}{BS}{Base Station}
\newacronym{bsr}{BSR}{Buffer Status Report}
\newacronym{bss}{BSS}{Business Support System}
\newacronym{ca}{CA}{Carrier Aggregation}
\newacronym{caas}{CaaS}{Connectivity-as-a-Service}
\newacronym{cav}{CAV}{Connected and Autonomous Vehicles}
\newacronym{cb}{CB}{Code Block}
\newacronym{cc}{CC}{Congestion Control}
\newacronym{ccid}{CCID}{Congestion Control ID}
\newacronym{cco}{CC}{Carrier Component}
\newacronym{cd}{CD}{Continuous Delivery}
\newacronym{cdd}{CDD}{Cyclic Delay Diversity}
\newacronym{cdf}{CDF}{Cumulative Distribution Function}
\newacronym{cdn}{CDN}{Content Distribution Network}
\newacronym{cli}{CLI}{Command-line Interface}
\newacronym{cn}{CN}{Core Network}
\newacronym{codel}{CoDel}{Controlled Delay Management}
\newacronym{comac}{COMAC}{Converged Multi-Access and Core}
\newacronym{cord}{CORD}{Central Office Re-architected as a Datacenter}
\newacronym{cornet}{CORNET}{COgnitive Radio NETwork}
\newacronym{cosmos}{COSMOS}{Cloud Enhanced Open Software Defined Mobile Wireless Testbed for City-Scale Deployment}
\newacronym{cots}{COTS}{Commercial Off-the-Shelf}
\newacronym{cp}{CP}{Control Plane}
\newacronym{cyp}{CP}{Cyclic Prefix}
\newacronym{up}{UP}{User Plane}
\newacronym{cpu}{CPU}{Central Processing Unit}
\newacronym{cqi}{CQI}{Channel Quality Information}
\newacronym{cr}{CR}{Cognitive Radio}
\newacronym{cran}{CRAN}{Cloud \gls{ran}}
\newacronym{crs}{CRS}{Cell Reference Signal}
\newacronym{csi}{CSI}{Channel State Information}
\newacronym{csirs}{CSI-RS}{Channel State Information - Reference Signal}
\newacronym{cu}{CU}{Central Unit}
\newacronym{d2tcp}{D$^2$TCP}{Deadline-aware Data center TCP}
\newacronym{d3}{D$^3$}{Deadline-Driven Delivery}
\newacronym{dac}{DAC}{Digital to Analog Converter}
\newacronym{dag}{DAG}{Directed Acyclic Graph}
\newacronym{das}{DAS}{Distributed Antenna System}
\newacronym{dash}{DASH}{Dynamic Adaptive Streaming over HTTP}
\newacronym{dc}{DC}{Dual Connectivity}
\newacronym{dccp}{DCCP}{Datagram Congestion Control Protocol}
\newacronym{dce}{DCE}{Direct Code Execution}
\newacronym{dci}{DCI}{Downlink Control Information}
\newacronym{dctcp}{DCTCP}{Data Center TCP}
\newacronym{dl}{DL}{Downlink}
\newacronym{dmr}{DMR}{Deadline Miss Ratio}
\newacronym{dmrs}{DMRS}{DeModulation Reference Signal}
\newacronym{drlcc}{DRL-CC}{Deep Reinforcement Learning Congestion Control}
\newacronym{drs}{DRS}{Discovery Reference Signal}
\newacronym{du}{DU}{Distributed Unit}
\newacronym{e2e}{E2E}{end-to-end}
\newacronym{earfcn}{EARFCN}{E-UTRA Absolute Radio Frequency Channel Number}
\newacronym{ecaas}{ECaaS}{Edge-Cloud-as-a-Service}
\newacronym{ecn}{ECN}{Explicit Congestion Notification}
\newacronym{edf}{EDF}{Earliest Deadline First}
\newacronym{embb}{eMBB}{Enhanced Mobile Broadband}
\newacronym{empower}{EMPOWER}{EMpowering transatlantic PlatfOrms for advanced WirEless Research}
\newacronym{enb}{eNB}{evolved Node Base}
\newacronym{endc}{EN-DC}{E-UTRAN-\gls{nr} \gls{dc}}
\newacronym{epc}{EPC}{Evolved Packet Core}
\newacronym{eps}{EPS}{Evolved Packet System}
\newacronym{es}{ES}{Edge Server}
\newacronym{etsi}{ETSI}{European Telecommunications Standards Institute}
\newacronym[firstplural=Estimated Times of Arrival (ETAs)]{eta}{ETA}{Estimated Time of Arrival}
\newacronym{eutran}{E-UTRAN}{Evolved Universal Terrestrial Access Network}
\newacronym{faas}{FaaS}{Function-as-a-Service}
\newacronym{fapi}{FAPI}{Functional Application Platform Interface}
\newacronym{fdd}{FDD}{Frequency Division Duplexing}
\newacronym{fdm}{FDM}{Frequency Division Multiplexing}
\newacronym{fdma}{FDMA}{Frequency Division Multiple Access}
\newacronym{fed4fire}{FED4FIRE+}{Federation 4 Future Internet Research and Experimentation Plus}
\newacronym{fir}{FIR}{Finite Impulse Response}
\newacronym{fit}{FIT}{Future \acrlong{iot}}
\newacronym{fpga}{FPGA}{Field Programmable Gate Array}
\newacronym{fr2}{FR2}{Frequency Range 2}
\newacronym{fs}{FS}{Fast Switching}
\newacronym{fscc}{FSCC}{Flow Sharing Congestion Control}
\newacronym{ftp}{FTP}{File Transfer Protocol}
\newacronym{fw}{FW}{Flow Window}
\newacronym{ge}{GE}{Gaussian Elimination}
\newacronym{gnb}{gNB}{Next Generation Node Base}
\newacronym{gop}{GOP}{Group of Pictures}
\newacronym{gpr}{GPR}{Gaussian Process Regressor}
\newacronym{gpu}{GPU}{Graphics Processing Unit}
\newacronym{gtp}{GTP}{GPRS Tunneling Protocol}
\newacronym{gtpc}{GTP-C}{GPRS Tunnelling Protocol Control Plane}
\newacronym{gtpu}{GTP-U}{GPRS Tunnelling Protocol User Plane}
\newacronym{gtpv2c}{GTPv2-C}{\gls{gtp} v2 - Control}
\newacronym{gw}{GW}{Gateway}
\newacronym{harq}{HARQ}{Hybrid Automatic Repeat reQuest}
\newacronym{hetnet}{HetNet}{Heterogeneous Network}
\newacronym{hh}{HH}{Hard Handover}
\newacronym{hol}{HOL}{Head-of-Line}
\newacronym{hqf}{HQF}{Highest-quality-first}
\newacronym{hss}{HSS}{Home Subscription Server}
\newacronym{http}{HTTP}{HyperText Transfer Protocol}
\newacronym{ia}{IA}{Initial Access}
\newacronym{iab}{IAB}{Integrated Access and Backhaul}
\newacronym{ic}{IC}{Incident Command}
\newacronym{ietf}{IETF}{Internet Engineering Task Force}
\newacronym{imsi}{IMSI}{International Mobile Subscriber Identity}
\newacronym{imt}{IMT}{International Mobile Telecommunication}
\newacronym{iot}{IoT}{Internet of Things}
\newacronym{ip}{IP}{Internet Protocol}
\newacronym{itu}{ITU}{International Telecommunication Union}
\newacronym{kpi}{KPI}{Key Performance Indicator}
\newacronym{kpm}{KPM}{Key Performance Measurement}
\newacronym{kvm}{KVM}{Kernel-based Virtual Machine}
\newacronym{los}{LoS}{Line of Sight}
\newacronym{lsm}{LSM}{Link-to-System Mapping}
\newacronym{lstm}{LSTM}{Long Short Term Memory}
\newacronym{lte}{LTE}{Long Term Evolution}
\newacronym{lxc}{LXC}{Linux Container}
\newacronym{m2m}{M2M}{Machine to Machine}
\newacronym{mac}{MAC}{Medium Access Control}
\newacronym{manet}{MANET}{Mobile Ad Hoc Network}
\newacronym{mano}{MANO}{Management and Orchestration}
\newacronym{mc}{MC}{Multi-Connectivity}
\newacronym{mcc}{MCC}{Mobile Cloud Computing}
\newacronym{mchem}{MCHEM}{Massive Channel Emulator}
\newacronym{mcs}{MCS}{Modulation and Coding Scheme}
\newacronym{mec2}{MEC}{Multi-access Edge Computing}
\newacronym{mec}{MEC}{Mobile Edge Computing}
\newacronym{mfc}{MFC}{Mobile Fog Computing}
\newacronym{mgen}{MGEN}{Multi-Generator}
\newacronym{mi}{MI}{Mutual Information}
\newacronym{mib}{MIB}{Master Information Block}
\newacronym{miesm}{MIESM}{Mutual Information Based Effective SINR}
\newacronym{mimo}{MIMO}{Multiple Input, Multiple Output}
\newacronym{ml}{ML}{Machine Learning}
\newacronym{mlr}{MLR}{Maximum-local-rate}
\newacronym[plural=\gls{mme}s,firstplural=Mobility Management Entities (MMEs)]{mme}{MME}{Mobility Management Entity}
\newacronym{mmtc}{mMTC}{Massive Machine-Type Communications}
\newacronym{mmwave}{mmWave}{millimeter wave}
\newacronym{mpdccp}{MP-DCCP}{Multipath Datagram Congestion Control Protocol}
\newacronym{mptcp}{MPTCP}{Multipath TCP}
\newacronym{mr}{MR}{Maximum Rate}
\newacronym{mrdc}{MR-DC}{Multi \gls{rat} \gls{dc}}
\newacronym{mse}{MSE}{Mean Square Error}
\newacronym{mss}{MSS}{Maximum Segment Size}
\newacronym{mt}{MT}{Mobile Termination}
\newacronym{mtd}{MTD}{Machine-Type Device}
\newacronym{mtu}{MTU}{Maximum Transmission Unit}
\newacronym{mumimo}{MU-MIMO}{Multi-user \gls{mimo}}
\newacronym{mvno}{MVNO}{Mobile Virtual Network Operator}
\newacronym{nalu}{NALU}{Network Abstraction Layer Unit}
\newacronym{nas}{NAS}{Network Attached Storage}
\newacronym{nat}{NAT}{Network Address Translation}
\newacronym{nbiot}{NB-IoT}{Narrow Band IoT}
\newacronym{nfv}{NFV}{Network Function Virtualization}
\newacronym{nfvi}{NFVI}{Network Function Virtualization Infrastructure}
\newacronym{ni}{NI}{Network Interfaces}
\newacronym{nic}{NIC}{Network Interface Card}
\newacronym{now}{NOW}{Non Overlapping Window}
\newacronym{nsm}{NSM}{Network Service Mesh}
\newacronym{nr}{NR}{New Radio}
\newacronym{nrf}{NRF}{Network Repository Function}
\newacronym{nsa}{NSA}{Non Stand Alone}
\newacronym{nse}{NSE}{Network Slicing Engine}
\newacronym{nssf}{NSSF}{Network Slice Selection Function}
\newacronym{o2i}{O2I}{Outdoor to Indoor}
\newacronym{oai}{OAI}{OpenAirInterface}
\newacronym{oaicn}{OAI-CN}{\gls{oai} \acrlong{cn}}
\newacronym{oairan}{OAI-RAN}{\acrlong{oai} \acrlong{ran}}
\newacronym{oam}{OAM}{Operations, Administration and Maintenance}
\newacronym{ofdm}{OFDM}{Orthogonal Frequency Division Multiplexing}
\newacronym{olia}{OLIA}{Opportunistic Linked Increase Algorithm}
\newacronym{omec}{OMEC}{Open Mobile Evolved Core}
\newacronym{onap}{ONAP}{Open Network Automation Platform}
\newacronym{onf}{ONF}{Open Networking Foundation}
\newacronym{onos}{ONOS}{Open Networking Operating System}
\newacronym{oom}{OOM}{\gls{onap} Operations Manager}
\newacronym{opnfv}{OPNFV}{Open Platform for \gls{nfv}}
\newacronym{oran}{O-RAN}{Open \gls{ran}}
\newacronym{orbit}{ORBIT}{Open-Access Research Testbed for Next-Generation Wireless Networks}
\newacronym{os}{OS}{Operating System}
\newacronym{oss}{OSS}{Operations Support System}
\newacronym{pa}{PA}{Position-aware}
\newacronym{pase}{PASE}{Prioritization, Arbitration, and Self-adjusting Endpoints}
\newacronym{pawr}{PAWR}{Platforms for Advanced Wireless Research}
\newacronym{pbch}{PBCH}{Physical Broadcast Channel}
\newacronym{pcef}{PCEF}{Policy and Charging Enforcement Function}
\newacronym{pcfich}{PCFICH}{Physical Control Format Indicator Channel}
\newacronym{pcrf}{PCRF}{Policy and Charging Rules Function}
\newacronym{pdcch}{PDCCH}{Physical Downlink Control Channel}
\newacronym{pdcp}{PDCP}{Packet Data Convergence Protocol}
\newacronym{pdsch}{PDSCH}{Physical Downlink Shared Channel}
\newacronym{pdu}{PDU}{Packet Data Unit}
\newacronym{pf}{PF}{Proportional Fair}
\newacronym{pgw}{PGW}{Packet Gateway}
\newacronym{phich}{PHICH}{Physical Hybrid ARQ Indicator Channel}
\newacronym{phy}{PHY}{Physical}
\newacronym{pmch}{PMCH}{Physical Multicast Channel}
\newacronym{pmi}{PMI}{Precoding Matrix Indicators}
\newacronym{powder}{POWDER}{Platform for Open Wireless Data-driven Experimental Research}
\newacronym{ppo}{PPO}{Proximal Policy Optimization}
\newacronym{ppp}{PPP}{Poisson Point Process}
\newacronym{prach}{PRACH}{Physical Random Access Channel}
\newacronym{prb}{PRB}{Physical Resource Block}
\newacronym{psnr}{PSNR}{Peak Signal to Noise Ratio}
\newacronym{pss}{PSS}{Primary Synchronization Signal}
\newacronym{pucch}{PUCCH}{Physical Uplink Control Channel}
\newacronym{pusch}{PUSCH}{Physical Uplink Shared Channel}
\newacronym{qam}{QAM}{Quadrature Amplitude Modulation}
\newacronym{qci}{QCI}{\gls{qos} Class Identifier}
\newacronym{qoe}{QoE}{Quality of Experience}
\newacronym{qos}{QoS}{Quality of Service}
\newacronym{quic}{QUIC}{Quick UDP Internet Connections}
\newacronym{ra}{RA}{Resouces Allocation}
\newacronym{rach}{RACH}{Random Access Channel}
\newacronym{ran}{RAN}{Radio Access Network}
\newacronym[firstplural=Radio Access Technologies (RATs)]{rat}{RAT}{Radio Access Technology}
\newacronym{rbg}{RBG}{Resource Block Group}
\newacronym{rcn}{RCN}{Research Coordination Network}
\newacronym{rc}{RC}{RAN Control}
\newacronym{rec}{REC}{Radio Edge Cloud}
\newacronym{red}{RED}{Random Early Detection}
\newacronym{renew}{RENEW}{Reconfigurable Eco-system for Next-generation End-to-end Wireless}
\newacronym{rf}{RF}{Radio Frequency}
\newacronym{rfc}{RFC}{Request for Comments}
\newacronym{rfr}{RFR}{Random Forest Regressor}
\newacronym{ric}{RIC}{\gls{ran} Intelligent Controller}
\newacronym{rlc}{RLC}{Radio Link Control}
\newacronym{rlf}{RLF}{Radio Link Failure}
\newacronym{rlnc}{RLNC}{Random Linear Network Coding}
\newacronym{rmr}{RMR}{RIC Message Router}
\newacronym{rmse}{RMSE}{Root Mean Squared Error}
\newacronym{rnis}{RNIS}{Radio Network Information Service}
\newacronym{rr}{RR}{Round Robin}
\newacronym{rrc}{RRC}{Radio Resource Control}
\newacronym{rrm}{RRM}{Radio Resource Management}
\newacronym{rru}{RRU}{Remote Radio Unit}
\newacronym{rs}{RS}{Remote Server}
\newacronym{rsrp}{RSRP}{Reference Signal Received Power}
\newacronym{rsrq}{RSRQ}{Reference Signal Received Quality}
\newacronym{rss}{RSS}{Received Signal Strength}
\newacronym{rssi}{RSSI}{Received Signal Strength Indicator}
\newacronym{rtt}{RTT}{Round Trip Time}
\newacronym{ru}{RU}{Radio Unit}
\newacronym{rw}{RW}{Receive Window}
\newacronym{rx}{RX}{Receiver}
\newacronym{s1ap}{S1AP}{S1 Application Protocol}
\newacronym{sa}{SA}{standalone}
\newacronym{sack}{SACK}{Selective Acknowledgment}
\newacronym{sap}{SAP}{Service Access Point}
\newacronym{sc2}{SC2}{Spectrum Collaboration Challenge}
\newacronym{scef}{SCEF}{Service Capability Exposure Function}
\newacronym{sch}{SCH}{Secondary Cell Handover}
\newacronym{scoot}{SCOOT}{Split Cycle Offset Optimization Technique}
\newacronym{sctp}{SCTP}{Stream Control Transmission Protocol}
\newacronym{sdap}{SDAP}{Service Data Adaptation Protocol}
\newacronym{sdk}{SDK}{Software Development Kit}
\newacronym{sdm}{SDM}{Space Division Multiplexing}
\newacronym{sdma}{SDMA}{Spatial Division Multiple Access}
\newacronym{sdn}{SDN}{Software-defined Networking}
\newacronym{sdr}{SDR}{Software-defined Radio}
\newacronym{seba}{SEBA}{SDN-Enabled Broadband Access}
\newacronym{sgsn}{SGSN}{Serving GPRS Support Node}
\newacronym{sgw}{SGW}{Service Gateway}
\newacronym{si}{SI}{Study Item}
\newacronym{sib}{SIB}{Secondary Information Block}
\newacronym{sinr}{SINR}{Signal to Interference plus Noise Ratio}
\newacronym{sip}{SIP}{Session Initiation Protocol}
\newacronym{siso}{SISO}{Single Input, Single Output}
\newacronym{sla}{SLA}{Service Level Agreement}
\newacronym{sm}{SM}{Service Model}
\newacronym{smo}{SMO}{Service Management and Orchestration}
\newacronym{smsgmsc}{SMS-GMSC}{\gls{sms}-Gateway}
\newacronym{snr}{SNR}{Signal-to-Noise-Ratio}
\newacronym{son}{SON}{Self-Organizing Network}
\newacronym{sptcp}{SPTCP}{Single Path TCP}
\newacronym{srb}{SRB}{Service Radio Bearer}
\newacronym{srn}{SRN}{Standard Radio Node}
\newacronym{srs}{SRS}{Sounding Reference Signal}
\newacronym{ss}{SS}{Synchronization Signal}
\newacronym{sss}{SSS}{Secondary Synchronization Signal}
\newacronym{st}{ST}{Spanning Tree}
\newacronym{svc}{SVC}{Scalable Video Coding}
\newacronym{tb}{TB}{Transport Block}
\newacronym{tcp}{TCP}{Transmission Control Protocol}
\newacronym{tdd}{TDD}{Time Division Duplexing}
\newacronym{tdm}{TDM}{Time Division Multiplexing}
\newacronym{tdma}{TDMA}{Time Division Multiple Access}
\newacronym{tfl}{TfL}{Transport for London}
\newacronym{tfrc}{TFRC}{TCP-Friendly Rate Control}
\newacronym{tft}{TFT}{Traffic Flow Template}
\newacronym{tgen}{TGEN}{Traffic Generator}
\newacronym{tip}{TIP}{Telecom Infra Project}
\newacronym{tm}{TM}{Transparent Mode}
\newacronym{to}{TO}{Telco Operator}
\newacronym{tr}{TR}{Technical Report}
\newacronym{trp}{TRP}{Transmitter Receiver Pair}
\newacronym{ts}{TS}{Technical Specification}
\newacronym{tti}{TTI}{Transmission Time Interval}
\newacronym{ttt}{TTT}{Time-to-Trigger}
\newacronym{tx}{TX}{Transmitter}
\newacronym{uas}{UAS}{Unmanned Aerial System}
\newacronym{uav}{UAV}{Unmanned Aerial Vehicle}
\newacronym{udm}{UDM}{Unified Data Management}
\newacronym{udp}{UDP}{User Datagram Protocol}
\newacronym{udr}{UDR}{Unified Data Repository}
\newacronym{ue}{UE}{User Equipment}
\newacronym{uhd}{UHD}{\gls{usrp} Hardware Driver}
\newacronym{ul}{UL}{Uplink}
\newacronym{um}{UM}{Unacknowledged Mode}
\newacronym{uml}{UML}{Unified Modeling Language}
\newacronym{upa}{UPA}{Uniform Planar Array}
\newacronym{upf}{UPF}{User Plane Function}
\newacronym{urllc}{URLLC}{Ultra Reliable and Low Latency Communications}
\newacronym{usa}{U.S.}{United States}
\newacronym{usim}{USIM}{Universal Subscriber Identity Module}
\newacronym{usrp}{USRP}{Universal Software Radio Peripheral}
\newacronym{utc}{UTC}{Urban Traffic Control}
\newacronym{vim}{VIM}{Virtualization Infrastructure Manager}
\newacronym{vm}{VM}{Virtual Machine}
\newacronym{vnf}{VNF}{Virtual Network Function}
\newacronym{volte}{VoLTE}{Voice over \gls{lte}}
\newacronym{voltha}{VOLTHA}{Virtual OLT HArdware Abstraction}
\newacronym{vr}{VR}{Virtual Reality}
\newacronym{vran}{vRAN}{Virtualized \gls{ran}}
\newacronym{vss}{VSS}{Video Streaming Server}
\newacronym{wbf}{WBF}{Wired Bias Function}
\newacronym{wf}{WF}{Waterfilling}
\newacronym{wg}{WG}{Working Group}
\newacronym{wlan}{WLAN}{Wireless Local Area Network}
\newacronym{osm}{OSM}{Open Source \gls{nfv} Management and Orchestration}
\newacronym{pnf}{PNF}{Physical Network Function}
\newacronym{drl}{DRL}{Deep Reinforcement Learning}
\newacronym{mtc}{MTC}{Machine-type Communications}
\newacronym{osc}{OSC}{O-RAN Software Community}
\newacronym{mns}{MnS}{Management Services}
\newacronym{ves}{VES}{\gls{vnf} Event Stream}
\newacronym{ei}{EI}{Enrichment Information}
\newacronym{fh}{FH}{Fronthaul}
\newacronym{fft}{FFT}{Fast Fourier Transform}
\newacronym{laa}{LAA}{Licensed-Assisted Access}
\newacronym{plfs}{PLFS}{Physical Layer Frequency Signals}
\newacronym{ptp}{PTP}{Precision Time Protocol}
\newacronym{lidar}{LiDAR}{Light Detection And Ranging}
\newacronym{dem}{DEM}{Digital Elevation Model}
\newacronym{dtm}{DEM}{Digital Terrain Model}
\newacronym{dsm}{DEM}{Digital Surface Models}
\newacronym{ota}{OTA}{Over-The-Air}
\newacronym{ns}{NS}{Network Slicing}
\newacronym{ne}{NE}{Nash Equilibrium}
\newacronym{hf}{HF}{High Frequency}
\newacronym{noma}{NOMA}{Non-Orthogonal Multiple Access}
\newacronym{sre}{SRE}{Smart Radio Environment}
\newacronym{ris}{RIS}{Reconfigurable Intelligent Surface}
\newacronym{inp}{InP}{Infrastructure Provider}
\newacronym{smf}{SMF}{Slicing Magangement Framework}
\newacronym{nsn}{NSN}{Network Slicing Negotiation}
\newacronym{sms}{SMS}{Slicing MAC Scheduler}
\newacronym{brd}{BRD}{Best Response Dynamics}
\newacronym{dssbr}{DSSBR}{Double Step Smoothed Best Response}
\newacronym{poa}{PoA}{Price of Anarchy}
\newacronym{pos}{PoS}{Price of Stability}
\newacronym{milp}{MILP}{Mixed Integer-Linear Program}
\newacronym{pod}{PoD}{Price of DSSBR}
\newacronym{roc}{ROC}{Radio Overload Control}
\newacronym{ciot}{cIoT}{critical Internet of Things}
\newacronym{embbpr}{eMBB Pr.}{enhanced Mobile BroadBand Premium}
\newacronym{embbbs}{eMBB Bs.}{enhanced Mobile BroadBand Basic}
\newacronym{en}{EN}{Edge Node}
\newacronym{ec}{EC}{Edge Computing}
\newacronym{sp}{SP}{Service Provider}
\newacronym{me}{ME}{Market Equilibrium}
\newacronym{so}{SO}{Social Optimum}
\newacronym{wso}{WSO}{Weighted Social Optimum}
\newacronym{ps}{PS}{Proportional Sharing}
\newacronym{eg}{EG}{Eisenberg-Gale program}
\newacronym{pe}{PE}{Pareto Efficiency}
\newacronym{nsw}{NSW}{Nash Social Welfare}
\newacronym{ef}{EF}{Envy-Freeness}
\newacronym{sub6}{sub6GHz}{Below 6GHz}
\newacronym{ncr}{NCR}{Network-Controlled Repeater}
\newacronym{nlos}{NLoS}{Non-Line of Sight}
\newacronym{src}{SRC}{Smart Radio Connection}
\newacronym{srd}{SRD}{Smart Radio Device}
\newacronym{cs}{CS}{Candidate Site}
\newacronym{tp}{TP}{Test Point}
\newacronym{fov}{FoV}{Field of View}
\newacronym{nrric}{near-RT RIC}{Near Real-time {RAN} Intelligent Controller}
\newacronym{e2ap}{E2AP}{E2 Application Protocol}
\newacronym{e2sm}{E2SM}{E2 Service Model}
\newacronym{nrtric}{non-RT RIC}{Non-Real-Time {RIC}}
\newacronym{itti}{ITTI}{Inter-task Interface}
\newacronym{bap}{BAP}{Backhaul Adaptation Protocol}
\newacronym{iabest}{IABEST}{Integrated Access and Backhaul Experimental large-Scale Tetbed}
\newacronym{teid}{TEID}{Tunnel Endpoint Identifier}
\newacronym{dlsch}{DL-SCH}{Downlink Shared Channel }
\newacronym{ulsch}{UL-SCH}{Uplink Shared Channel }
\newacronym{rsu}{RSU}{Road Side Unit}
\newacronym{v2v}{V2V}{Vehicle-to-Vehicle}
\newacronym{SNR}{SNR}{Signal-to-Noise Ration}

\usepackage{textcomp}
\usepackage{lineno}
\usepackage{booktabs}
\usepackage{balance} 
\usepackage{tabularx}
\usepackage{mathtools}
\usepackage{graphicx}
\usepackage{graphics}
\usepackage{epstopdf}
\usepackage{amsmath,amssymb,amsfonts}
\usepackage[english]{babel}
\usepackage{amsthm}
\usepackage{soul}
\usepackage{color}
\usepackage{subcaption}
\usepackage{caption}
\usepackage{wrapfig}
\usepackage{comment}
\usepackage{float}
\usepackage{amssymb}
\usepackage{xcolor}
\usepackage{algorithm}
\usepackage{algpseudocode}

\def\BibTeX{{\rm B\kern-.05em{\sc i\kern-.025em b}\kern-.08em
    T\kern-.1667em\lower.7ex\hbox{E}\kern-.125emX}}
\begin{document}

\title{Open RAN-empowered V2X Architecture:\\ Challenges, Opportunities, and Research Directions}

\author{
\IEEEauthorblockN{Francesco Linsalata, Eugenio Moro, Maurizio Magarini, Umberto Spagnolini, Antonio Capone}

\IEEEauthorblockA{Dipartimento di Elettronica, Informazione e Bioingegneria, Politecnico di Milano, Milan, Italy} 

E-mails: \{name.surname\}@polimi.it}

\maketitle

\begin{abstract}
Advances in the automotive industry and the ever-increasing demand for Connected and Autonomous Vehicles (CAVs) are pushing for a new epoch of networked wireless systems. Vehicular communications, or Vehicle-to-Everything (V2X), are expected to be among the main actors of the future beyond 5G and 6G networks. However, the challenging application requirements, the fast variability of the vehicular environment, and the harsh propagation conditions of high frequencies call for sophisticated control mechanisms to ensure the success of such a disruptive technology. While traditional Radio Access Networks (RAN) lack the flexibility to support the required control primitives, the emergent concept of Open RAN (O-RAN) appears as an ideal enabler of V2X communication orchestration. However, effectively integrating the two ecosystems is still an open issue. This paper discusses possible integration strategies, highlighting the challenges and opportunities of leveraging O-RAN to enable real-time V2X control. Additionally, we enrich our discussion with potential research directions stemming from the current state-of-the-art, and we provide preliminary simulation results that validate the effectiveness of the proposed integration.
\end{abstract}


\section{Introduction}

Connected and autonomous vehicles (CAVs) will revolutionize transportation systems. CAVs will guarantee safer travel, less pollution, and optimized solutions in terms of time and costs with respect to the old branded cars.
In this context, connectivity plays a crucial role, providing enablers such as network infrastructure, network distribution, cloud-to-edge resources, localization, data technologies, and governance. The research community is focusing on new directions that may be relevant in future beyond 5G (B5G) and 6G wireless networks, including applications for the automotive sectors, as 5G moves closer to being widely deployed around the world \cite{6GV2Xchallenges}.

Currently, dedicated short-range communications (DSRC) and cellular vehicle-to-everything (C-V2X) are the two main V2X communication strategies. The former is supported by the IEEE 802.11p standard, whereas the latter is promoted by the Long Term Evolution (LTE) or New Radio (NR) standards. As suggested by the latest 3rd Generation Partnership Project (3GPP) recommendations, vehicular communications at high frequencies, e.g., millimetre-waves (mmWaves) or sub-THz, are the key 6G technology that will make V2X systems  possible~\cite{rel17, tutorialBoban}.
However, those frequencies are not very reliable, particularly in dynamic and prone to link line-of-sight (LoS) blockage environments \cite{LinsalataLoSmap}.

Owing to highly dynamic environments, V2X communications experience rapidly varying link conditions, which can greatly affect communication performance. 
Moreover, the issue of maintaining large-scale connectivity in a highly dynamic environment is the most challenging aspect to address in 6G V2X systems. The high dimensionality of the data, high data rate, stringent latency requirements, challenges with data harvesting and user privacy protection, the dynamic nature of numerous operator parameters and vendor-proprietary data, as well as irrational bandwidth requirements for training, are a few of them \cite{6GV2Xchallenges}.
The 6G V2X is an extremely challenging context that calls for sophisticated orchestration, control, and optimization solutions to ensure seamless service provisioning in V2X. However, the required data collection and control primitives go beyond what is currently possible in traditional \gls{ran} deployments. 

\gls{oran} is an emerging architectural overhaul for mobile radio networks that have recently gained extreme popularity, both in the academic and industrial contexts~\cite{OranAliance,polese2022understanding}. Through open interfaces, \gls{bs} disaggregation, and centralized control loop, \gls{oran} promises to introduce flexibility and programmatic control in the current and future generations of cellular networks. As such, it represents the ideal candidate to unlock the potential of V2X through large-scale data collection and dynamic control. 

\Gls{oran} concepts have been applied to optimize several endeavors of traditional \glspl{ran} deployments~\cite{lacava2022programmable}, with some solutions being also tested in vehicular communication scenarios~\cite{mollahasani2021}. However, the research effort concerning the potential synergies of \gls{oran} and V2X is still unexplored. As such, the challenges of successfully integrating the two technologies are still open, and the opportunities are still to be fully exploited. 

In this visionary paper, we discuss the opportunity of leveraging on \gls{oran} as an open platform to enable dynamic control of vehicular communications. In Sec. II, we will first focus on the challenges of integrating the two architectures by proposing possible strategies. Then, we give a series of research directions to harness the large-scale control capabilities enabled by the aforementioned integration. In Sec. III, we provide some simulation results to show the benefits \gls{oran}-empowered V2X solutions. Lastly, Sec. IV concludes the work.

%The remainder of this paper is organized as follows. Section~\ref{sec}
\section{Integration Challenges and Opportunities}
\label{sec:integration}
\begin{figure} [t!]
    \centering
\subfloat[\label{fig:bs}]{ \includegraphics[width=\columnwidth]{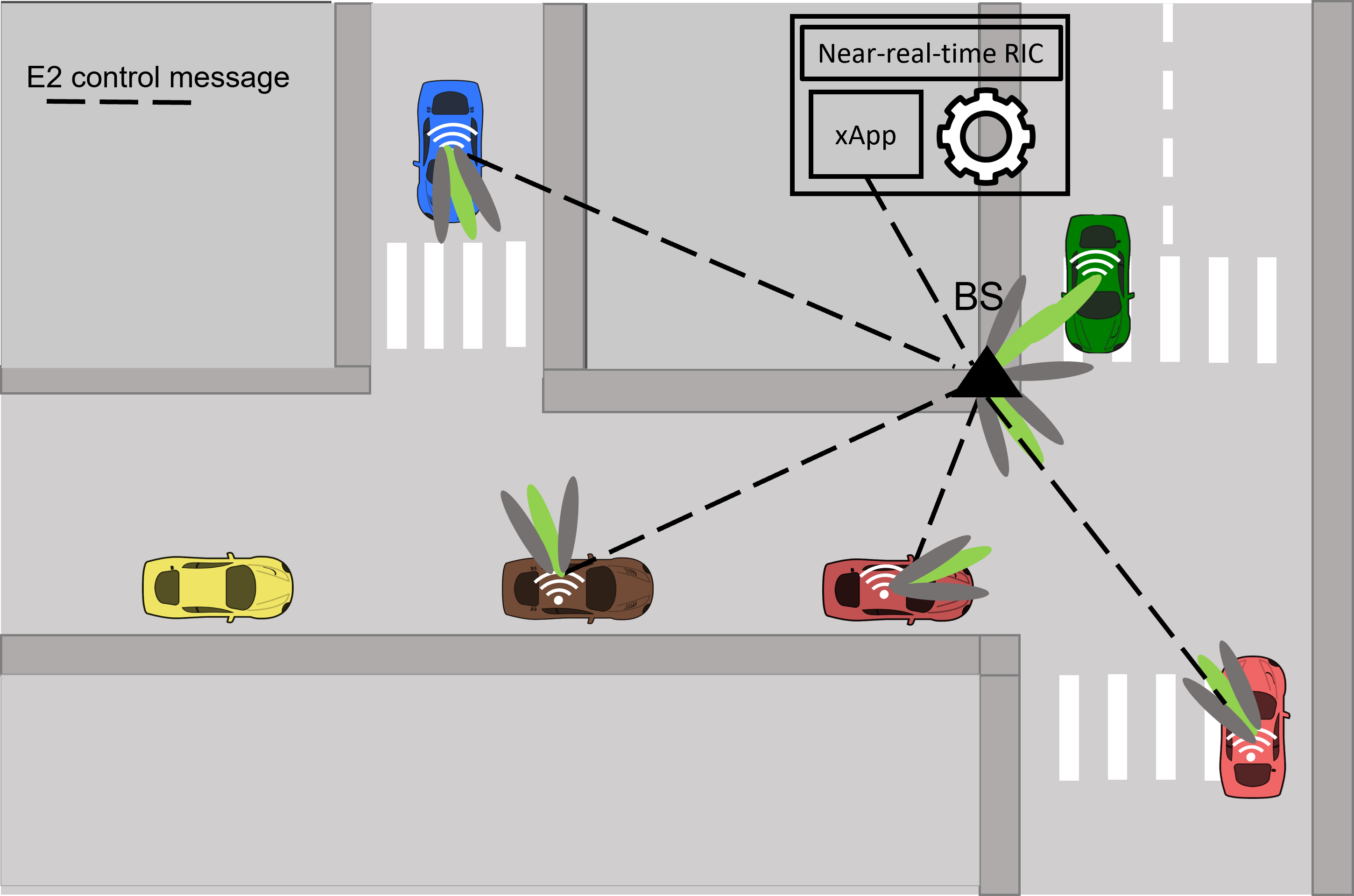}}

    \subfloat[ \label{fig:bs}]{\includegraphics[width=.99\columnwidth]{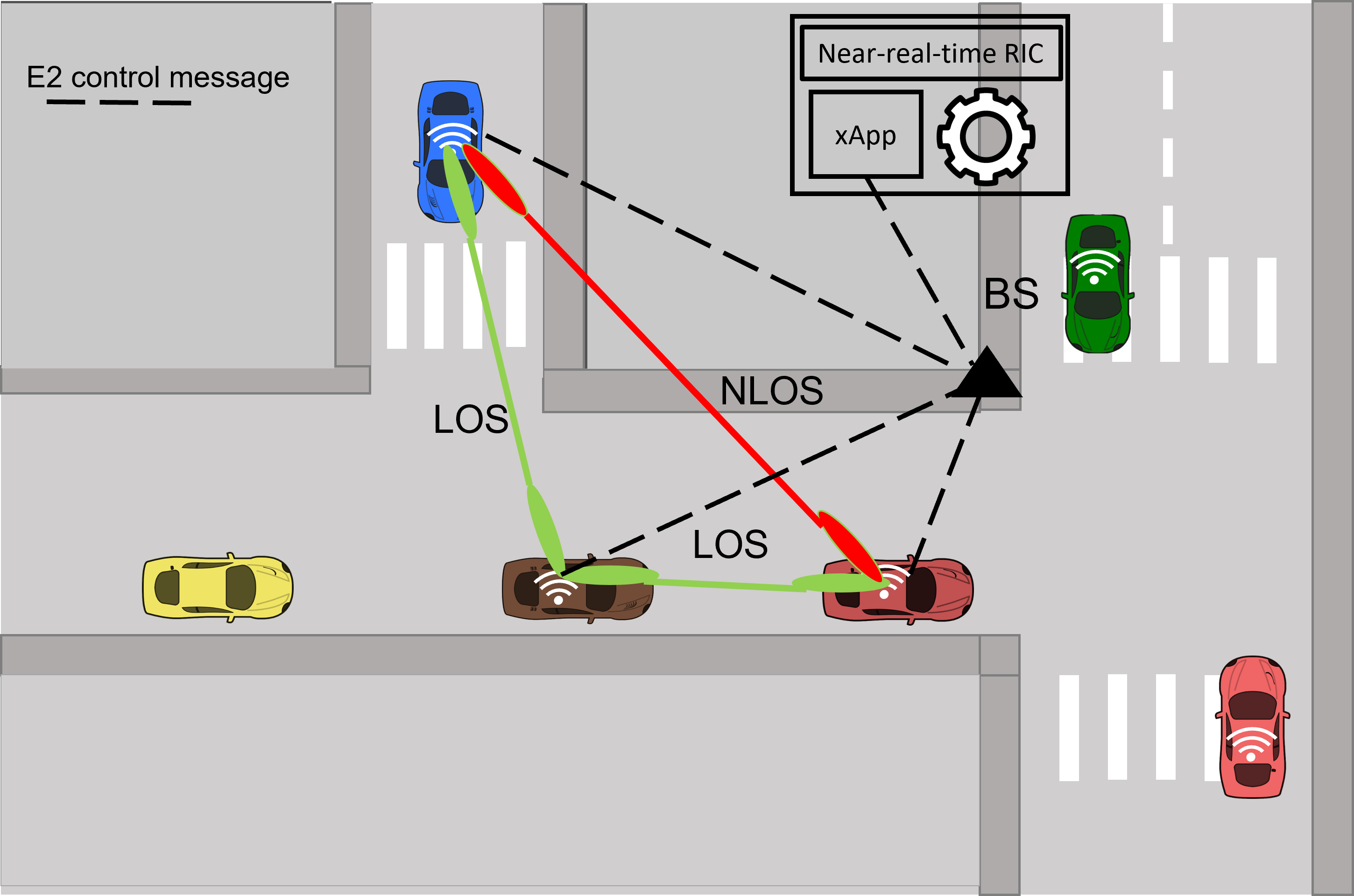}}    
    \caption{ORAN-empowered (a) beam selection and (b) relay allocation}
    \label{fig:usecases}
\end{figure}
The \gls{oran} architecture features the possibility of applying centralized control to the \gls{ran} through the so-called \glspl{ric}. These functional components can implement arbitrary data collection and control logic by communicating with the network infrastructure (i.e. \glspl{bs}) thanks to open and standardized interfaces. In particular, \gls{oran} introduced a \gls{nrric}, which operates on a $1\,$ms to $1\,$s time scale and is capable of operating under stringent latency requirements. Arbitrary data collection and control mechanisms are then implemented through the so-called xApps: network applications that run on the primitives exposed by the \gls{nrric}.  Additionally, \gls{oran} has also standardized the \gls{nrtric}: a centralized control loop operating on a slower time scale but with broader network visibility. As such, it enables large-scale orchestration and policing mechanisms implemented as network applications called rApps. When applied to V2X, these two control loops can potentially unlock significant optimization and orchestration gains with respect to the current architecture. 

\subsection{V2X and O-RAN Integration}

Due to the peculiarities of the V2X system, some modifications to the \gls{oran} architecture are required. 
The \glspl{ric} communicates with the network infrastructure through open interfaces: the E2 interface for the \gls{nrtric} and the O1 and O2 interfaces for the \gls{nrric}. The E2 interface is functionally split into two protocols: \gls{e2ap} - enabling communication between the \gls{ran} component and the \gls{nrric} - and \gls{e2sm} - which defines the control semantics as \glspl{sm}. By standard, a \gls{bs} is equipped with E2 terminations to enable data collection and control. In V2X systems, an E2 termination could also be included in \glspl{rsu}, both to enable their dynamic control and to tap into the wealth of CAV-related information that they make available. E2 messages can thus be multiplexed together with the other communications in the \gls{rsu} control plane. Additionally, there is a case to be made for including an E2 termination in the CAVs themselves, as it will be motivated in the following paragraphs and showed in Figure \ref{fig:usecases}. However, \gls{e2ap} does not currently support mobility and proper modifications to the protocol are needed before the CAV can be directly accessed by the xApps deployed on the \gls{nrric}. In both cases, new \gls{sm} definitions will also be required to support data collection and control applied to V2X. 

The \gls{nrtric} employs the O1 interface to apply \gls{smo} functions over the entire network infrastructure and the O2 interface to control the life-cycle of network components. These interfaces will require modifications that are naturally similar to the E2 case. In particular, the O1 interface will require the definition of dedicated V2X Management Services or at least the modification of existing ones. Furthermore, inserting O2 terminations in the \glspl{rsu} could enable adaptive network deployment strategies that can selectively activate/deactivate all the network components of V2X systems to scale the system performance when required and decrease energy consumption and interference when it is possible. 

\subsection{Research Directions}

In the following, we describe four fundamental open challenges of V2X, highlighting how these can be successfully addressed through \gls{oran}-based solutions. At the same time, we discuss the key architectural modifications required in each case. 

\textbf{Resources Allocation.}
Efficient radio resource allocation mechanisms in V2X are required to guarantee robustness against the harsh V2X propagation conditions \cite{ShinRA2023}. This is especially relevant for direct vehicle-to-vehicle (V2V) connections, also known as sidelink. In this case, there is the need to choose the optimal time-frequency resource to be allocated to allow direct communication between the CAVs. According to the standard, a central entity (i.e., a \gls{bs} or a \gls{rsu}) is expected to allocate the radio resources. This mechanism is hindered by the limited perception of the central entity with respect to each V2V link condition and traffic requirement \cite{ShinRA2023}. In this context, an xApp could gather data about the vehicle's position and mobility, as well as channel status and interference profile. This information can be processed to adapt the allocation strategies to the fast-varying V2V environment. As previously mentioned, the \gls{oran} architecture could be extended to include E2 terminations directly in the CAVs. In this case, the data collection and control procedures could happen directly to the vehicles involved in the communication, unlocking the possibility of extremely fine and precise tuning of the allocation mechanism. 

\textbf{Beam selection and management.} Beam-based communications are necessary for the high frequencies employed in V2X. Due to the high vehicular mobility, traditional beam selection and management mechanisms are considered inadequate, and sophisticated solutions are required instead. Data-driven approaches are effective in providing fast beam alignment, and \gls{oran} represents an ideal enabler for these solutions. For instance, an xApp could process relevant information coming from the urban layout, vehicle positioning, and past successful beam alignment to produce an up-to-date probabilistic codebook and deploy them to the base stations \cite{mizmizi2022fastening}. The same information can be used to train machine learning-based beam steering mechanisms \cite{BeamAlignmentML}. \gls{oran} has been specifically designed to facilitate such data-driven control loops. However, currently, standardized \gls{sm} do not specifically support the collection of the required data types.

\textbf{Relay Assignment.}
Propagation at high frequencies is subject to severe attenuation and mostly requires direct link visibility conditions. This is a key issue in dynamic environments such as vehicular ones, where frequent link misalignment and blockage occasions can easily occur. The resulting non-LoS (NLoS) condition leads to severe system performance degradation and, consequently, intermittent connectivity.
Link blockage can be mitigated by relaying and multi-hop mechanisms, for example, exploiting nearby CAVs, RSU, or smart environments technologies as Intelligent Reflecting Surfaces (IRS)~\cite{moro2021planning}.
However, the limited relaying resources capabilities and the fast changes in the network make the relay assignment a significant challenge in the current V2X systems. In this context, an \gls{oran}-based solution is capable of determining optimal relay assignment by leveraging on the data collection and information fusion capabilities of the \gls{nrric}. Similarly to what has been discussed before, this context makes a case for integrating an E2 termination in the CAVs. As a result of this integration, the relay assignment choices are delivered directly to the involved CAVs, speeding up the process and reducing the signaling overhead. 
An introductory case study on this challenge will be presented in the next section.

\textbf{V2X Network Digital Twin.}
The 6G V2X communications will exploit the cooperation among CAVs to augment environment perception and to enable the creation of a digital replica of the surrounding environments \cite{ding2022digital}.
To obtain an accurate real-time digital reproduction of the physical environment, the envisioned digital twin-enabled V2X system has to use high-definition 3D maps and combine multi-modal sensory data from several vehicles' onboard sensor data.
The acquisition of data from the global navigation satellite system (GNSS), cameras, lidars, and radars distributed over multiple road entities must be appropriately orchestrated over a fast 6G V2X RAN. Once again, the \gls{oran} architecture is well-positioned to take on this orchestration role. Thanks to the fast processing capabilities of the \gls{nrric}, the large data volume involved can be gathered, filtered, and pre-processed in parallel over separate network sections. The separated data streams can then be aggregated into the \gls{nrtric}, where the digital twin is built and kept up-to-date with the fast-changing state of the physical infrastructure. rApps deployed on the \gls{nrtric} can access the digital twin to produce tightly optimized policies based on a complete and precise vision of the entire physical system.
\section{A first simulation analysis}
\begin{table}[t!] 
\footnotesize
\centering
\caption{{Simulation variables}} \label{table:simulation_parameters}
%\footnotesize
\begin{tabular}{ | c | c | p{42mm} |}
\hline 
\multicolumn{2}{|c|}{Simulation parameters} \\
\hline
Central Frequency & 28 GHz\\
Max EIRP & 23 dBm  \\
Simulation time & 300 s \\
Urban Pathloss model & 3GPP and ITU  \cite{LinsalataLoSmap} \\
Vehicular traffic density &  50-70 veh/km \\
Max hops number &  4 \\
%Relay scenarios & SNR, distance and no relay \\
%Pathloss model & ThreeGppUmiStreetCanyon \\
%Channel model type & V2V-Urban  \\  
\hline 
	\end{tabular}
\end{table}

\begin{figure} [t!]
    \centering
  {\includegraphics[width=\columnwidth]{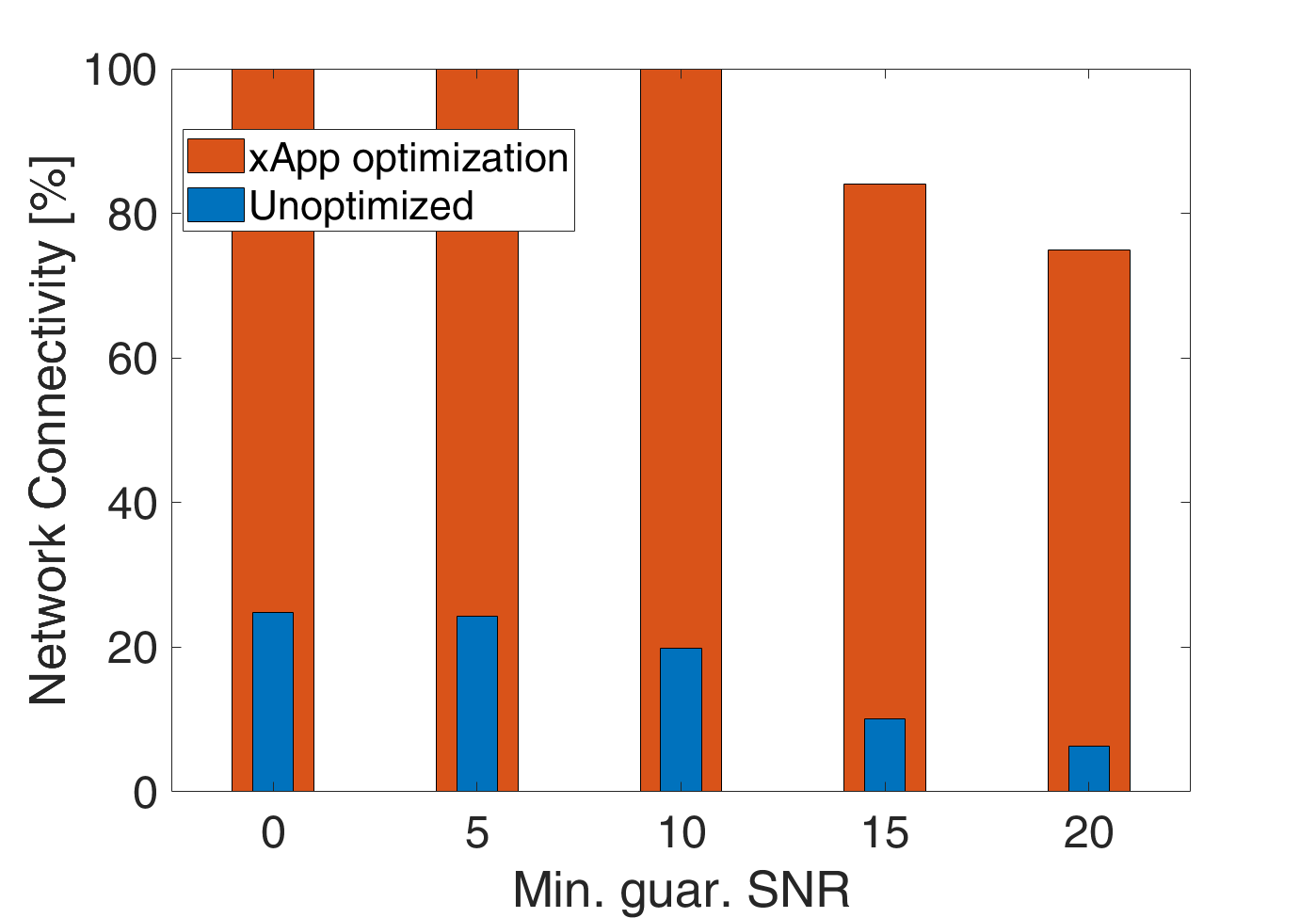}}    
    \caption{Network connectivity versus different SNR thresholds.}
    \label{fig:connectivity_barplot_vnc}
\end{figure}

\begin{figure} [t!]
    \centering
  {\includegraphics[width=\columnwidth]{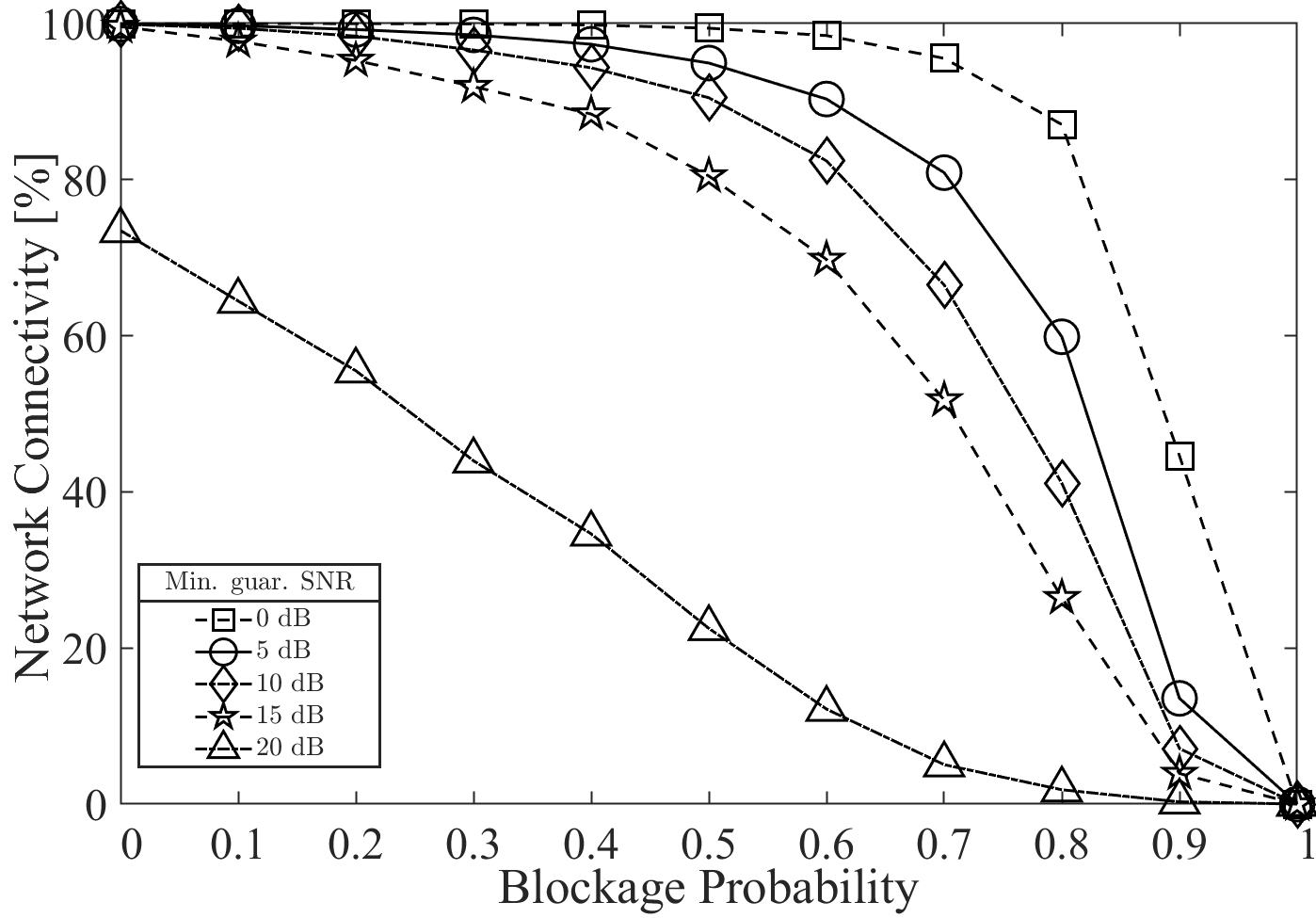}}    
    \caption{
    Network connectivity versus blockage probability for different SNR thresholds.}
    \label{fig:pb_vs_connections}
\end{figure}

\label{sec:results}
In this section, we conduct a preliminary case study based on a typical vehicular communication scenario to demonstrate the effectiveness of the proposed system. We simulated multiple CAVs traversing an urban intersection. 
However, \gls{v2v} links can be blocked by other moving road users or buildings. \Glspl{rsu} and other \glspl{cav} can act as relays when \gls{los} is obstructed. 

We replicated a V2X urban scenario using a vehicular channel simulator \cite{LinsalataLoSmap} with the parameters in Tab. \ref{table:simulation_parameters}. We emulated the xApp's behaviour by calculating alternative routes between vehicle pairs using a minimum \gls{SNR} as a constraint. Such a constraint stands for the minimum \gls{qos} required for different \gls{v2v} use cases \cite{5GAA}. 

We use the network connectivity as a performance evaluation metric for assessing the robustness of the V2X network. It is intricately linked to the density of CAVs and the reliability of the vehicle mesh configuration.
A robust \gls{v2v} network connectivity is required to exchange safety-critical information throughout the entire navigation area. 

Results in Figure~\ref{fig:connectivity_barplot_vnc} demonstrate that the proposed xApp-empowered system has the potential of guaranteeing high vehicular connectivity even for high levels of minimum guaranteed \gls{snr}. The unoptimized baseline approach, which considers only the direct links, shows that no more than 25\% of the vehicles can establish a connection throughout the simulation time window. 
The slight degradation observed at high \gls{snr} thresholds primarily stems from the limited availability of links that meet the required criteria.

Figure \ref{fig:pb_vs_connections} shows the impact of vehicular blockage on the V2X network and how the xApp reacts to different SNR conditions. We conducted multiple experiments by forcing the higher number of blockers and the minimum \gls{SNR} to assess the xApp's ability to choose a path according to predefined communication performance criteria, even in very challenging conditions. Indeed, a higher number of blockers corresponds to an increased probability of encountering NLOS links, characterized by lower \gls{snr} values. Consequently, the presence of blockers limits the available relay options due to the reduced signal quality, complicating relay selection processes.
In environments with dense blocker distributions, the challenges in identifying suitable relays become pronounced, necessitating effective relay selection methods capable of operating effectively under adverse propagation conditions. 

\section{Concluding Remarks}
As the world moves towards a more connected and automated future, the need for reliable and efficient communication between vehicles and network infrastructure has become increasingly important. 
This paper focused on using Open Radio Access Network (O-RAN) architecture for V2X communication. We debated how the O-RAN architecture has the potential to provide a more flexible, scalable, and efficient solution compared to current V2X systems. Moreover, we discussed novel research directions and justified them through introductory results, which will be subject to further analysis and validation.

\section*{Acknowledgment}
\small{
This work was supported by the European Union under the Italian National Recovery and Resilience Plan (NRRP) of NextGenerationEU, partnership on “Telecommunications of the Future” (PE00000001 - program “RESTART”, Structural Project 6GWINET).
}
\bibliographystyle{IEEEtran}
\bibliography{biblio.bib}

\begin{thebibliography}{10}
\providecommand{\url}[1]{#1}
\csname url@samestyle\endcsname
\providecommand{\newblock}{\relax}
\providecommand{\bibinfo}[2]{#2}
\providecommand{\BIBentrySTDinterwordspacing}{\spaceskip=0pt\relax}
\providecommand{\BIBentryALTinterwordstretchfactor}{4}
\providecommand{\BIBentryALTinterwordspacing}{\spaceskip=\fontdimen2\font plus
\BIBentryALTinterwordstretchfactor\fontdimen3\font minus
  \fontdimen4\font\relax}
\providecommand{\BIBforeignlanguage}[2]{{%
\expandafter\ifx\csname l@#1\endcsname\relax
\typeout{** WARNING: IEEEtran.bst: No hyphenation pattern has been}%
\typeout{** loaded for the language `#1'. Using the pattern for}%
\typeout{** the default language instead.}%
\else
\language=\csname l@#1\endcsname
\fi
#2}}
\providecommand{\BIBdecl}{\relax}
\BIBdecl

\bibitem{6GV2Xchallenges}
M.~Noor-A-Rahim, Z.~Liu, H.~Lee, M.~O. Khyam, J.~He, D.~Pesch, K.~Moessner,
  W.~Saad, and H.~V. Poor, ``6g for vehicle-to-everything (v2x) communications:
  Enabling technologies, challenges, and opportunities,'' \emph{Proceedings of
  the IEEE}, vol. 110, no.~6, pp. 712--734, 2022.

\bibitem{rel17}
\BIBentryALTinterwordspacing
3rd Generation Partnership~Project. (2021) {3GPP Release 17}. [Online].
  Available: \url{https://www.3gpp.org/release-17}
\BIBentrySTDinterwordspacing

\bibitem{tutorialBoban}
M.~H.~C. Garcia, A.~Molina-Galan, M.~Boban, J.~Gozalvez, B.~Coll-Perales,
  T.~Şahin, and A.~Kousaridas, ``A tutorial on {5G NR V2X} communications,''
  \emph{IEEE Communications Surveys \& Tutorials}, vol.~23, no.~3, pp.
  1972--2026, 2021.

\bibitem{LinsalataLoSmap}
F.~Linsalata, S.~Mura, M.~Mizmizi, M.~Magarini, P.~Wang, M.~N. Khormuji,
  A.~Perotti, and U.~Spagnolini, ``{LoS-Map Construction for Proactive Relay of
  Opportunity Selection in 6G V2X Systems},'' \emph{IEEE Transactions on
  Vehicular Technology}, pp. 1--15, 2022.

\bibitem{OranAliance}
``{0-RAN alliance},'' \url{https://www.o-ran.org/}.

\bibitem{polese2022understanding}
M.~Polese, L.~Bonati, S.~D’Oro, S.~Basagni, and T.~Melodia, ``{Understanding
  O-RAN: Architecture, Interfaces, Algorithms, Security, and Research
  Challenges},'' \emph{IEEE Communications Surveys \& Tutorials}, pp. 1--1,
  2023.

\bibitem{lacava2022programmable}
A.~Lacava, M.~Polese, R.~Sivaraj, R.~Soundrarajan, B.~S. Bhati, T.~Singh,
  T.~Zugno, F.~Cuomo, and T.~Melodia, ``Programmable and customized
  intelligence for traffic steering in 5g networks using open ran
  architectures,'' \emph{arXiv preprint arXiv:2209.14171}, 2022.

\bibitem{mollahasani2021}
S.~Mollahasani, M.~Erol-Kantarci, and R.~Wilson, ``Dynamic cu-du selection for
  resource allocation in o-ran using actor-critic learning,'' in \emph{2021
  IEEE Global Communications Conference (GLOBECOM)}, 2021, pp. 1--6.

\bibitem{ShinRA2023}
C.~Shin, E.~Farag, H.~Ryu, M.~Zhou, and Y.~Kim, ``Vehicle-to-everything (v2x)
  evolution from 4g to 5g in 3gpp: Focusing on resource allocation aspects,''
  \emph{IEEE Access}, vol.~11, pp. 18\,689--18\,703, 2023.

\bibitem{mizmizi2022fastening}
M.~Mizmizi, F.~Linsalata, M.~Brambilla, F.~Morandi, K.~Dong, M.~Magarini,
  M.~Nicoli, M.~N. Khormuji, P.~Wang, R.~A. Pitaval \emph{et~al.}, ``Fastening
  the initial access in {5G NR} sidelink for {6G} {V2X} networks,''
  \emph{Vehicular Communications}, vol.~33, p. 100402, 2022.

\bibitem{BeamAlignmentML}
S.~Rezaie, C.~N. Manchón, and E.~de~Carvalho, ``Location- and
  orientation-aided millimeter wave beam selection using deep learning,'' in
  \emph{ICC 2020 - 2020 IEEE International Conference on Communications (ICC)},
  2020, pp. 1--6.

\bibitem{moro2021planning}
E.~Moro, I.~Filippini, A.~Capone, and D.~De~Donno, ``Planning mm-wave access
  networks with reconfigurable intelligent surfaces,'' in \emph{2021 IEEE 32nd
  Annual International Symposium on Personal, Indoor and Mobile Radio
  Communications (PIMRC)}, 2021, pp. 1401--1407.

\bibitem{ding2022digital}
C.~Ding and I.~W.-H. Ho, ``{Digital-Twin-enabled City-model-aware Deep Learning
  for Dynamic Channel Estimation in Urban Vehicular Environments},'' \emph{IEEE
  Transactions on Green Communications and Networking}, vol.~6, no.~3, pp.
  1604--1612, 2022.

\bibitem{5GAA}
\BIBentryALTinterwordspacing
S.~Conway, ``V2x technology benchmark testing,'' 5th Generation Automotive
  Association, Tech. Rep. ET 13-49, Sep. 2018. [Online]. Available:
  \url{https://www.fcc.gov/ecfs/filing/109271050222769}
\BIBentrySTDinterwordspacing

\end{thebibliography}

\end{document}